# The influence of alloying on the stacking fault energy of gold from density functional theory calculations

Anuj Goyal[1,†], Yangzhong Li (李扬中) [1,††], Aleksandr Chernatynskiy[2], Jay. S. Jayashankar[3,†††], Michael C. Kautzky[3], Susan B. Sinnott[4] and Simon R. Phillpot[1,*]

[1] Department of Materials Science and Engineering, University of Florida, Gainesville FL 32611 USA

[2] Department of Physics, Missouri University of Science and Technology, Rolla MO 65409, USA

[3] Seagate Technology. 7801 Computer Ave. S, Bloomington MN 55435 USA

[4] Department of Materials Science and Engineering, The Pennsylvania State University, State College PA, 16802 USA

Abstract:

The generalized stacking fault (SFE) energy curves of pure gold (Au) and its binary alloys with transition metals are determined from density functional theory (DFT). Alloy elements Ag, Al, Cu, Ni, Ti, Zr, Zn, In, Ga, Sn, Mn, Cd, Sn, Ta and Cr are substituted into Au at concentrations up to 4%. A comparison of various proposed methodologies to calculate SFEs is given. The intrinsic SFE decreases for all alloying elements from its value for pure Au, but SFE energies (both stable and unstable) vary strongly with the distance of the alloying element from the stacking fault region, and with alloy concentration. The compositional dependence of the SFE on the volume change associated with alloying element is determined. This work demonstrates that the SFE is strongly influenced by misfit strain caused by the alloying elements. Moreover, the computed generalized SFE curves provide information valuable to developing an understanding of the deformation behavior of Au and Au-alloys.

[†] Current Address: National Renewable Energy Laboratory, Golden, CO 80401, USA

†† Current Address: Sino-French Institute of Nuclear Engineering and Technology, Sun Yat-sen University, Zhuhai, Guangdong, China

††† Current Address: Cardiovascular Systems Inc., 1225 Old Highway 8 NW, St. Paul, MN 55112

* Corresponding Author: sphil@mse.ufl.edu

## 1. Introduction

Gold (Au) is in high demand as a conductor in microelectronic applications[1–3] due to its high electrical conductivity and good chemical stability. However, its mechanical integrity in devices applications is questionable as Au is mechanically soft. Improving the strength of gold either by alloying or via metallurgical processes such as work hardening, and precipitation hardening has been attempted, albeit with limited success[4–7]. This provides the motivation to investigate ways of improving the strength of pure gold without significantly sacrificing the electrical properties; here, therefore we examine the effects of alloying in the dilute limit. Mechanical deformation in face centered cubic (FCC) metal such as Au takes place primarily via dislocation motion[8–10]. Stacking faults (SFs) are known to govern the activation of various dislocation processes[11–15]. Of particular relevance to deformation behavior is the generalized stacking fault energy (GSFE) curve, which seems to be strongly correlated with the deformation mechanisms[12,16], dislocation nucleation from a crack tip[11,14,15], ideal shear strength of single crystals[17], twins[18] and other deformation processes, particularly as the microstructure size shrinks to the nanoscale regime[19].

FCC metals such as Pt, Pd, Al, Ni, have high stacking fault energies (SFEs) energies (greater than 125 mJ/m$^2$), Cu has an intermediate energy of 35 - 60 mJ/m$^2$, while Au and Ag lie in the low SFE regime (less than 35 mJ/m$^2$)[20–27]. Experimental studies[21,22] report a rather wide range of values (32 - 50 mJ/m$^2$) for the SFE of gold. However, based on recent measurements[28] using high-resolution transmission electron microscopy (HRTEM) and first principles based SFE calculations[24], the true value likely lies in at the lower end of this range or even just below: 25 - 35 mJ/m$^2$. The wide variation in experimental values is a result of the experimental complexity, with many factors needing to be well controlled in order to obtain accurate values[20,21].

For low stacking fault energy metal such as Au, dislocations extend into two Shockley partials on either side of an intrinsic stacking fault. Rice[11] showed that the energy barrier to nucleate the trailing partial depends on the unstable stacking fault energy[11]. Therefore in FCC metals like Ni and Al which have a lower $(\gamma_U - \gamma_I)$ energy barrier, it is relatively easy to nucleate the trailing partial. They, thus, have a relatively smaller stacking fault width compared to FCC metals like Cu, Au and Ag. Furthermore, indentation experiments[8,29,30] and complementary atomic-scale simulations[31,32] on Au thin films have clearly shown that the onset of plastic deformation involves emission and glide of Shockley partial dislocations.

More recently, combined experimental and simulation studies on Co-Ni[33], Ni-Al[34] and Mg-based[35,36] alloy systems have shown segregation of alloying elements at the stacking fault layer, resulting in very low stacking fault energies and the formation of stacking fault regions in these alloys systems that are much larger than in the pure metal. Therefore, segregation of solute atoms to stacking faults, referred to in literature as Suzuki segregation[37], is an important concept that quantifies the energetics of interaction of alloying element with the stacking fault region, and helps provide insights into the mechanical behavior of alloy systems. Based on these considerations and recent studies[16,24,38–41] it is evident that an accurate description of the GSFE curve is necessary to understand deformation behavior in FCC metals.

However, there is very limited analysis of SFE in Au alloys and, because it has an atypically low SFE, one has to be careful in applying trends seen in other FCC metal alloys directly to Au alloys. Here, therefore, we use first-principles electronic-structure calculations at the level of density functional theory (DFT) to determine the SFE of Au alloyed with other FCC, body centered cubic (BCC), and hexagonal close packed (HCP) metals in dilute amounts (with a maximum concentration of 4%). All of the alloying metals considered are cheaper than Au and more easily available. We find that intrinsic SFE value for pure Au decreases for all alloying elements and varies strongly with the distance of the alloying element from the stacking fault region. We demonstrate that the compositional dependence of the SFE is strongly influenced by misfit strain and the volume change associated with the alloying elements. Our simulations in dilute Au alloys provide insights into strengthening and also identify the possibilities of future studies.

The body of the manuscript is organized as follows. Computational methods to calculate general stacking fault energy (GSFE) curve and first-principle calculations are discussed in Section 2. In Section 3, the computed stacking fault energies results for pure Au and its alloys are presented in detail. These results are discussed in Section 4. Our conclusions are in Section 5.

**2. Methodology**

2.1    Atomistic simulations of stacking fault

The generalized stacking fault energy is a measure of the energy penalty for shearing two adjacent atomic planes. Starting from a perfect crystal, as two neighboring atomic planes glide past one another the energy per unit area of slip plane reaches a maximum at the unstable stacking fault, with energy $\gamma_U$ and then decreases to a local energy minimum at a relatively stable configuration; this is the stable or intrinsic stacking fault, with energy $\gamma_I$. This energy-displacement curve, often known as the generalized stacking fault energy (GSFE) curve or gamma curve, originally introduced by Vitek[42,43], is a powerful theoretical concept, although it cannot be mapped experimentally except at the stable stacking fault itself. Literature on shear deformation in FCC metals indicates that slip on {111} planes in the <110> directions is the major operative slip system.[24,44–46] An ideal FCC lattice consists of an ABCABCABC stacking sequence of close packed {111} planes (see Fig. 1A); <110>{111} slip involves, for example, a layer A moving over a C layer along <110>, as shown in Fig. 1B. Slip of atoms in layer A (top layer) over the atoms in the C layer along <110> direction leads to a steeper increase in energy than if the A atoms are shifted first to B positions. Such a displacement of atoms in layer A over atoms in layer C leads to an ABC|BCABC sequence, with the vertical bar denoting the location of the stacking fault. The stacking fault region is bounded by partial dislocations formed due to shear displacements and therefore its formation is often described in terms of a Burgers vector equation, given, as splitting of a Burgers vector $\frac{1}{2}[\bar{1}10]$ of a full dislocation into Burgers vectors $\frac{1}{6}[\bar{2}11]$ and $\frac{1}{6}[\bar{1}2\bar{1}]$, corresponding to the two Shockley partial dislocations (see Fig. 1B). For the {111} slip plane in a FCC structure such as Au, the stable stacking fault configuration corresponds to a slip of $a_0/\sqrt{6}$ in the $\langle 11\bar{2}\rangle$ direction for atoms in layer A over atoms in layer C.

For calculations of the stacking fault energy over the (111) plane, the smallest computational supercell is composed of an orthorhombic unit with lattice vectors oriented along $[11\bar{2}]$, $[\bar{1}10]$ and $[111]$ directions of the conventional FCC unit, with lengths $a_0\sqrt{6}/2$, $a_0/\sqrt{2}$ and $a_0\sqrt{3}$ (where, $a_0$ is the lattice parameter of FCC unit) respectively. The unit cell contains six atoms on three close packed {111} planes, in the ABC stacking sequence (Fig. 1A).

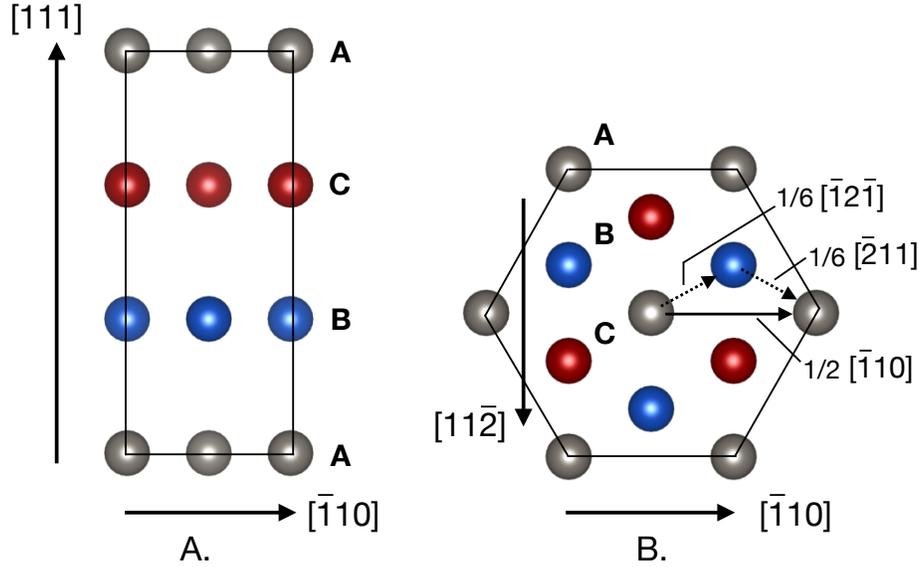

Figure 1: A. The geometry of the orthorhombic computational cell with three lattice vectors parallel to the $[11\bar{2}]$, $[\bar{1}10]$ and $[111]$ directions of the conventional fcc cubic unit cell. Atoms within this cell occupy the {111} close packed planes with ABC stacking along $[111]$. B. Schematic showing atoms in the A layer sliding over the atoms in C layer along the perfect dislocation $\frac{1}{2}[\bar{1}10]$, which splits into two Shockley partial dislocations $\frac{1}{6}[\bar{2}11]$ + $\frac{1}{6}[\bar{1}2\bar{1}]$. Atoms in the A, B and C layers are shown in grey, blue and red, respectively.

There are a number of different schemes to calculate the stacking fault energy using atomistic simulations. These can be broadly classified into, (i) the slab deformation approach[47–49], (ii) the shear deformation approach[17,25,46], and (iii) the direct estimation of energy difference between the perfect structure and structure with a stacking fault[50].

(i)   Slab deformation approach

In this method, to generate the stacking fault along $\{111\}[11\bar{2}]$, the supercell is divided into two halves with the lower half remaining fixed and the upper half rigidly displaced in the $[11\bar{2}]$ direction in small increments to gradually generate a stacking fault along $\{111\}[11\bar{2}]$. The lattice vectors of the simulation supercell are oriented along the $[11\bar{2}]$, $[\bar{1}10]$ and $[111]$ directions. After each incremental displacement, all the atoms are allowed to relax along $[111]$ direction, which is perpendicular to the shear plane. Periodic boundary conditions are applied in the $[11\bar{2}]$ and $[\bar{1}10]$ directions only and a vacuum of 20 Å is created on each side of the (111) faces. A schematic of the slab deformation method is shown in Fig. 2A.

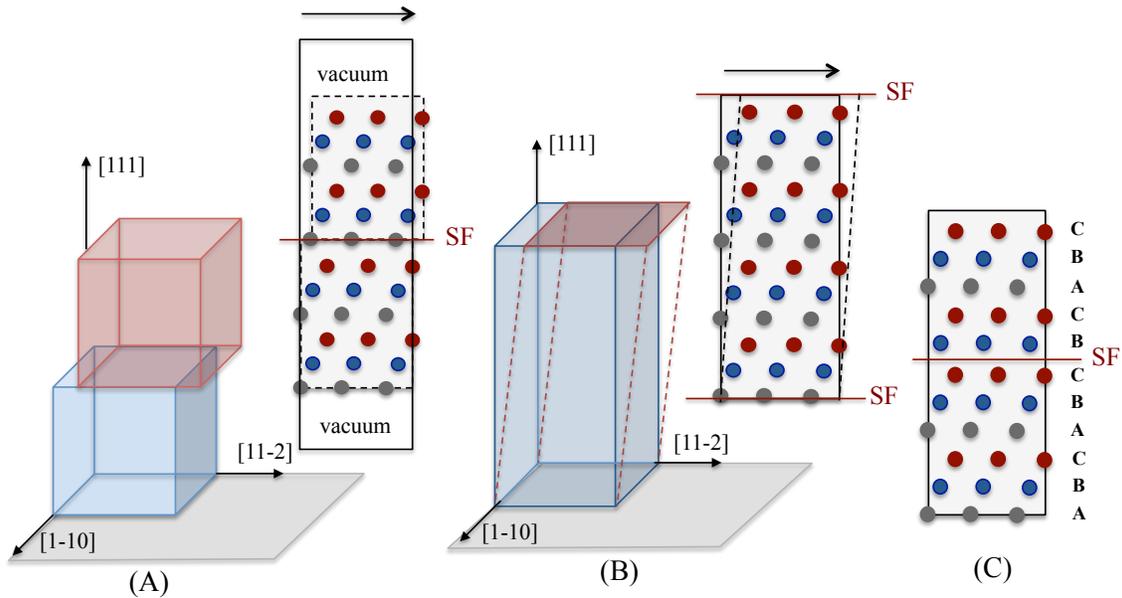

Figure 2: Schematic of the approaches used to calculate stacking fault energy. (A) Slab deformation approach, (B) Shear deformation approach, (C) Direct approach. Both (A) and (B) approaches can be used to obtain full GSFE curves, whereas approach (C) can only obtain the stable stacking fault energy. (111) Planes with stacking ABC along [111] direction are shown in grey, blue and red atoms, with solid red line representing the location of stacking fault.

(ii)   Shear deformation approach

This approach is also referred to in the literature as the simple alias shear method[17,46]. Atoms in the top (111) plane are displaced in the shear direction. (see Fig. 2B). The atoms in all the {111} planes are then allowed to relax in the direction normal to the shear plane, i.e. in the [111] direction while relaxation in the in-plane directions $[1\bar{1}0]$ and $[11\bar{2}]$ is prohibited to prevent sheared atoms from returning to their original non-sheared configuration. To obtain shear of the (111) plane along the $\langle 11\bar{2}\rangle$ direction in an orthorhombic computational cell with lattice vectors oriented along the $[11\bar{2}]$, $[\bar{1}10]$ and [111] directions, a deformation matrix **D** is applied. The deformed lattice vectors $\bar{R}$ of the orthorhombic cell after the alias shear along the $[11\bar{2}]$ direction, are then obtained by:

$$\bar{R} = DR, \tag{1}$$

where,

$$D = \begin{bmatrix} 1 & 0 & \varepsilon \\ 0 & 1 & 0 \\ 0 & 0 & 1 \end{bmatrix}. \tag{2}$$

Here, R represents the initial lattice vectors of orthorhombic cell and D is the deformation matrix with $\varepsilon$ corresponding to the shear strain. To generate a stable stacking fault along $\{111\}[11\bar{2}]$, $\varepsilon = \sqrt{2}/6$ is required if the length of orthorhombic c-axis (which is along [111] direction of conventional fcc lattice) is $a_0\sqrt{3}$. This corresponds to a displacement of $a_0\sqrt{6}/6$ of the top {111} layer along $[11\bar{2}]$ as shown in Fig. 1B.

(iii)   Direct approach

To obtain an intrinsic stacking fault, a layer A of atoms is removed resulting in a break in the conventional ABCABCABC stacking sequence, and resulting in formation of ABC|BCABC sequence, a shown in Fig. 2C. The energy of intrinsic stacking fault, can then be simply calculated from the difference of the energy of faulted and perfect supercell, as

$$\Delta E = (E_{faulted} - E_{perfect})n, \tag{3}$$

where $E_{faulted}$ and $E_{perfect}$ are the energies per unit atom of the system with and without the stacking fault, respectively and n is the total number of atoms in the system with the stacking fault. The stacking fault energy is then defined as the energy per unit area A of the stacking fault region

$$\gamma_{SF} = \frac{\Delta E}{A}. \qquad (4)$$

Equations 3 and 4 are also used with both slab and shear deformation approaches to obtain the general stacking fault energy (GSFE) curves, with $E_{faulted}$ and $E_{perfect}$ the energies per unit atom for the sheared and non-sheared (i.e., perfect) structures, respectively.

The direct approach only yields the intrinsic stacking fault energy; to generate the full GSFE curve, either slab or shear approaches are necessary. The shear approach allows the use of smaller system size, i.e., its volume is about half of the slab approach, which clearly reduces the computational cost and gives very accurate results for stacking fault energies, as discussed in the results section. However, when constructing fault structures with different displacements along the shear direction, the cell shape changes using the shear deformation approach and hence, it requires higher energy cutoff and denser k-meshes to ensure accurate convergence compared to the slab approach, which has fixed shape and dimensions.

## 2.2  DFT calculations

DFT based first-principles calculations of stacking faults are performed with the Vienna Ab-Initio Simulation Package (VASP)[51–53] using the projector-augmented-wave (PAW) method[54,55]. The electron exchange and correlation potentials are tested with both the local density approximation (LDA) and Perdew-Burke-Ernzerhof (PBE) method. The calculated lattice parameter $a_0$, elastic constant matrix $C_{ij}$, bulk modulus B, and shear modulus G, for LDA and PBE functionals for exchange-correlations are compared with available experimental values in the literature in Table 1, from which it is clear that the LDA approximation yields better agreement with the experimental values than PBE for pure Au.

A 2x2x2 supercell with 48 atoms is used for the DFT calculations of stacking fault curves in pure Au. This supercell contains 6 {111} layers with 8 atoms in each layer. The Brillouin zone (BZ) is sampled using a 6 x 10 x 4 mesh, constructed according to the Monkhorst-Pack scheme. The integration over the BZ uses the Methfessel-Paxton smearing method with 0.2 eV smearing width; a cutoff energy of 400 eV for the plane waves is used for all the calculations. The convergence criterion for the energy difference is $10^{-6}$ eV and the Hellman-Feynman force components acting on atoms are relaxed to at least $10^{-3}$ eV Å$^{-1}$. To compute the SFE in alloy systems, we have employed three difference supercell sizes – 24 atoms (1x2x2) (with basic unit as 6 atom orthorhombic cell), 48 atoms (2x2x2) and 96 atoms (2x4x2) with the approximate overall alloy concentrations of 4% (Au$_{23}$X), 2% (Au$_{47}$X) and 1 % (Au$_{95}$X), respectively. The shape and size of the alloy supercell is first relaxed and then SFE calculations are performed on this initial relaxed structure using alias shear method with fix volume and size.

Table 1: The values of lattice parameter $a_0$ in Å; single crystal elastic constants C$_{ij}$, bulk modulus B and shear modulus G in GPa for pure Au from DFT calculations using PAW potentials in VASP (LDA and PBE for exchange correlations functional). These results are compared with previous DFT and experimental values.

| Method | $a_0$ | $C_{11}$ | $C_{12}$ | $C_{44}$ | B | G |
|---|---|---|---|---|---|---|
| EXPT[56]. | 4.078 | 201.6 | 169.7 | 45.4 | 180.3 | 27.0 |
| LDA (this work) | 4.052 | 217.1 | 181.3 | 39.7 | 193.2 | 30.9 |
| PBE (this work) | 4.160 | 150.1 | 132.1 | 21.4 | 138.1 | 16.4 |
| LDA[24,57] | 4.06 | 211.1 | 184.1 | 38.9 | 193.1 | 28.7 |
| | | 202.1 | 174.2 | 37.9 | | |
| PBE[24,58] | 4.164 | 159.1 | 136.7 | 27.6 | 137.6 | 18.8 |
| | 4.17 | 149.6 | 130.9 | 25.1 | 137.1 | |

## 3. Results

### 3.1 Stacking fault energy of pure Au

In this section we report the GSFE curves for pure Au from DFT calculations and validate our results against experimentally available values. The stable $\gamma_I$ and the unstable $\gamma_U$ SFEs for the (111) [11$\bar{2}$] shear deformation are computed from the GSFE curves as shown in Fig. 3. The Burgers vector along [11$\bar{2}$], $b_P^{\langle 11\bar{2} \rangle} = \frac{1}{6}[11\bar{2}]a_0$ has a length of $a_0/\sqrt{6}$. The unstable $\gamma_U$ and stable $\gamma_I$ stacking faults are located at the shear displacements of $b_P^{\langle 11\bar{2} \rangle}/2$ and $b_P^{\langle 11\bar{2} \rangle}$, respectively. The calculated SFE values for pure Au are listed in Table 2 and they compare very well with the experimental values. The reported GSFE curves using LDA and GGA-PBE exchange-correlation functionals differ slightly; this can be attributed to the difference in lattice parameter and elastic constants (Table 1) predicted by these two functionals. The values from LDA agree better with the experimental literature, as do the elastic constants and lattice parameters. We also compared the stacking fault energies from the three different methods discussed in Sec. 2. The three methods agree very well (within 3%) as summarized in Table 2. Hence, for our subsequent calculations on pure Au and for dilute alloy concentrations, we use the LDA in our DFT calculations, except where specifically mentioned.

Table 2: Stable and the unstable stacking fault energies on the GSFE curve for Au calculated using three different stacking fault methods from DFT calculations and compared with experimental (stable stacking fault energies) and previous simulation data. The reported values for SFE are for 12 (111) atomic planes between the stacking fault layers with atomic relaxations.

| Method | $\gamma_{ISFE}$ (mJ/m$^2$) | | $\gamma_{USFE}$ (mJ/m$^2$) | |
|---|---|---|---|---|
| | LDA | PBE | LDA | PBE |
| Alias Shear | 31.3 | 26.5 | 97.0 | 76.4 |
| Slab Method | 32.5 | 30.1 | 100.0 | 69.4 |
| Direct Method | 33.2 | | 101.5 | |
| DFT literature[24] | | 27.9 | | 66.5 |
| Expt. | | 23[59], 30 –35[28,60,61] | | |

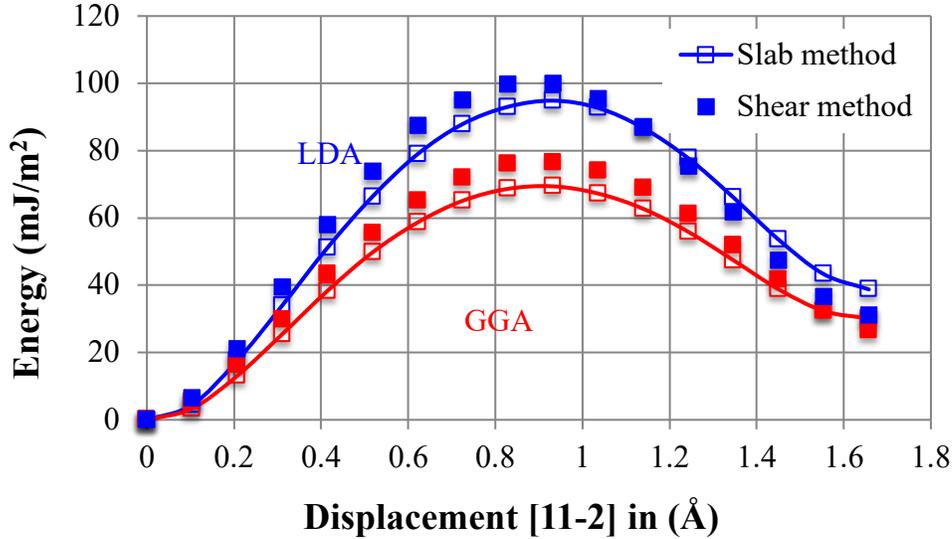

Figure 3: The general stacking fault energy (GSFE) curve for pure Au, calculated on the 1x1x2 supercell using the DFT calculations. Solid data points correspond to calculations done using the alias shear deformation method and the open data points + curve represents the values from the slab method.

Both atomic relaxation and the spacing between the stacking fault layers affect the SFE computed from theoretical DFT based simulations[62–66]. We find that for Au, the effect of atomic relaxation along [1-10] and [111] normal to the glide direction [11-2] is larger for the unstable SFE than for the stable SFE: the stable and unstable SFEs are lower by about 6% and 21%, respectively, due to atomic relaxations. Our results for changes in the SFE due to atomic relaxations are similar to those reported earlier for Au and other FCC systems[46,63]. To understand the sensitivity of atomic relaxations on the GSFE for alloy systems, we performed additional calculations in Au-Cu system. As shown in Fig. 4, we find that allowing for additional in-plane (marked as rlx-XZ) relaxation along [1-10] and full (rlx-XYZ) relaxations lowers the stable and unstable SFE lower by 1% and 10%, respectively, compared to the case where atomic relaxations are allowed only normal to the glide plane (111) (rlx-Z) and along the glide direction [11-2] (rlx-YZ). The full and additional relaxations in [1-10] direction allow for local atomic shuffling in the glide plane, thereby, providing the system a path to lower its energy compared to the constrained atomic relaxations along [11-2] and [111]. The local shuffling of atoms under additional relaxation disrupts the smooth energy versus displacement profile of the GSFE curve and similar effects has also been observed in previous computational studies on other FCC and HCP systems[46,64,67].

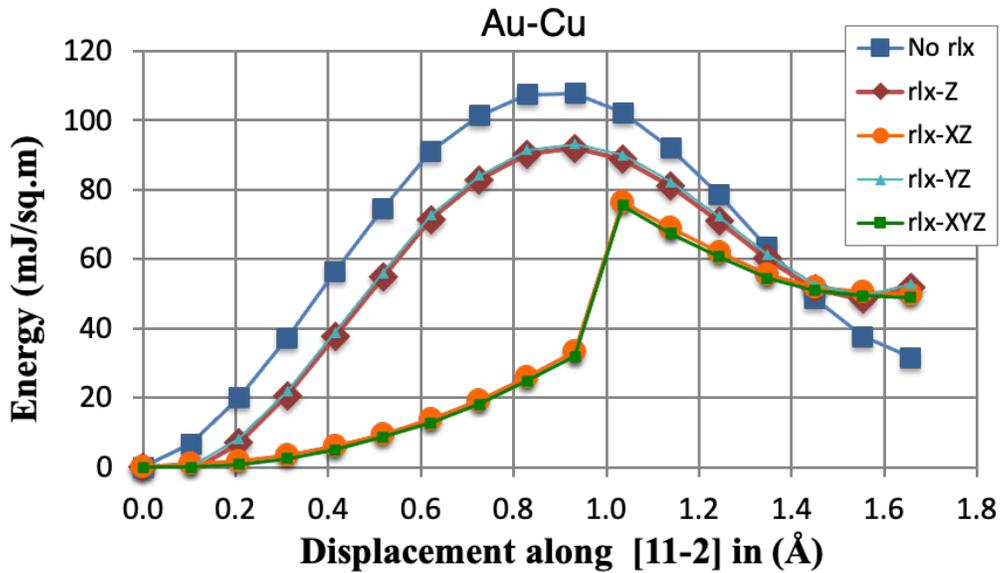

Figure 4: The general stacking fault energy curves for Au-Cu (2%) alloy as function of atomic relaxations obtained using alias shear method. Directions X, Y and Z correspond to [1-10], [11-2] and [111], respectively.

In addition, the spacing between the stacking fault layers must be large enough that the atoms in atomic planes between the stacking faults feel an environment similar to that of the perfect bulk. We investigated the influence of the number of layers (# = 3, 6, 9, 12 and 18 layers) between the stacking faults on the SFE values. We find that the difference both stable and unstable SFE change by less than 3% as the number of layers increases from 6 to 18; this is consistent with similar analyses in other FCC systems[63,65,66]. This relative independence on system size is associated with the short range of the strain field associated with the stacking fault[49,66,68,69].

Based on the results of this technical study, we choose a spacing of 6 atom planes between stacking faults in our calculations and have used the simple alias shear deformation method, with atomic positions relaxed normal to the glide direction for the alloy calculations.

### 3.2.1 SFE of Dilute alloys

The alloying atom (X) is substituted at a Au site ($X_{Au}^{\times}$) in a typical (111) layer. Given the periodic boundary conditions, we effectively have a stacking fault layer of infinite extent in the (111) plane and a periodic repeating stacking fault along [111] direction. The alloying atom can be placed in any of the six (111) layers including the stacking fault layer, as shown in Fig. 5A. We compute both the stable and unstable stacking fault energy for different alloying elements (FCC – Ag, Al, Cu, Ni; BCC – Cr, Ta; HCP – Ti, Zr, Zn, Cd and other crystal structure types – Mn, Ga, In and Sn) as function of specific position of substitution site in a (111) layer from the stacking fault. SFE values for the alloy systems strongly depend on distance, d, of the substitution site from the stacking fault layer, as shown in Fig. 5B, with different trends for different elements. For example, the stable stacking fault energy $\gamma_I$ in Au$_{47}$Ni and Au$_{47}$Ti systems increases significantly as the Ni and Ti atoms are substituted close to the stacking fault region, whereas Au$_{47}$Al shows a strong decrease in the $\gamma_I$. The change in energy of the system due to interaction of the alloying element with the stacking fault is in principle analogous to the interaction of impurities with defects like dislocations and grain boundaries. We explore this effect in the discussion section and also correlate change in SFE with position of the alloying element from the SF layer with the volume change and the size mismatch between the alloying element and Au atom.

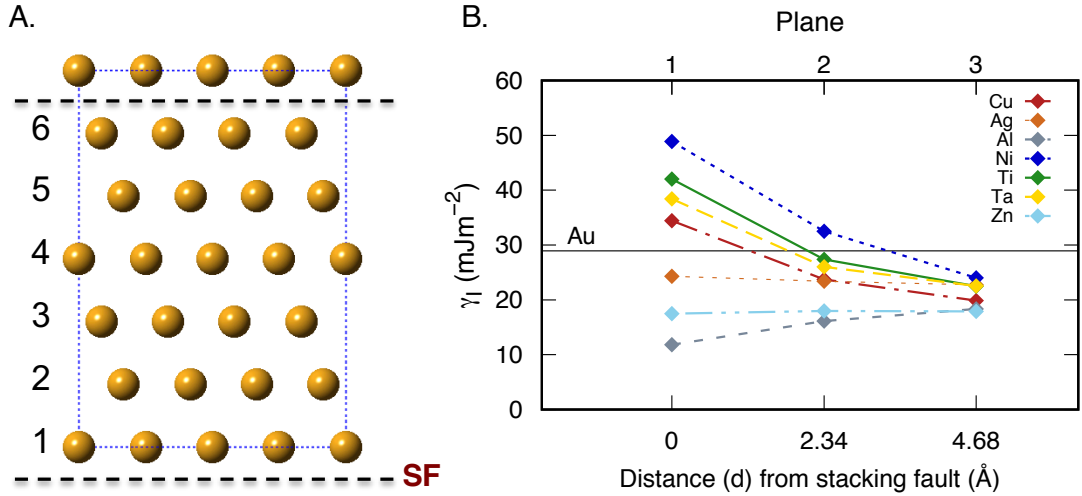

Figure 5: Intrinsic stacking fault energy of alloying atom (Cu, Ag, Al, Ni, Ti, Ta and Zn) as function of distance from the stacking fault region for Au$_{47}$X system (2% alloy concentration). The alloying (or impurity) atom is substituted on a Au site from plane #1 to 6 at a time in the stacking fault structure.

In addition to the variations in stacking fault energies with the position of the substitution site, the energies also vary significantly with the impurity type. Figure 6 shows the energy as a function of the relaxation volume ΔV of the alloying element for an overall impurity concentration of 2%. The relaxation volume is the difference in the volume between the DFT supercell with (Au$_{47}$X) and without (Au$_{48}$) the alloying element and it is function of defect concentration. All impurities result in a decrease in both $\gamma_I$ and $\gamma_U$ with respect to the energies of pure Au, when substituted in the (111) layer farthest away (layer #3 and #4) from the stacking fault layer. The decrease in $\gamma_I$ is in the range 0 – 12 mJ/m$^2$, with Al and Zn having the largest reductions, and Mn and Zr having the lowest reductions. However, when the alloying atom is substituted in the stacking fault layer, the trends in both stable and unstable SFE change. Elements Ti, Cr, Ni, Ta, Mn and Zr show an increase in SFE with respect to pure Au while Al, In, Cd, Sn and Ga show a decrease.

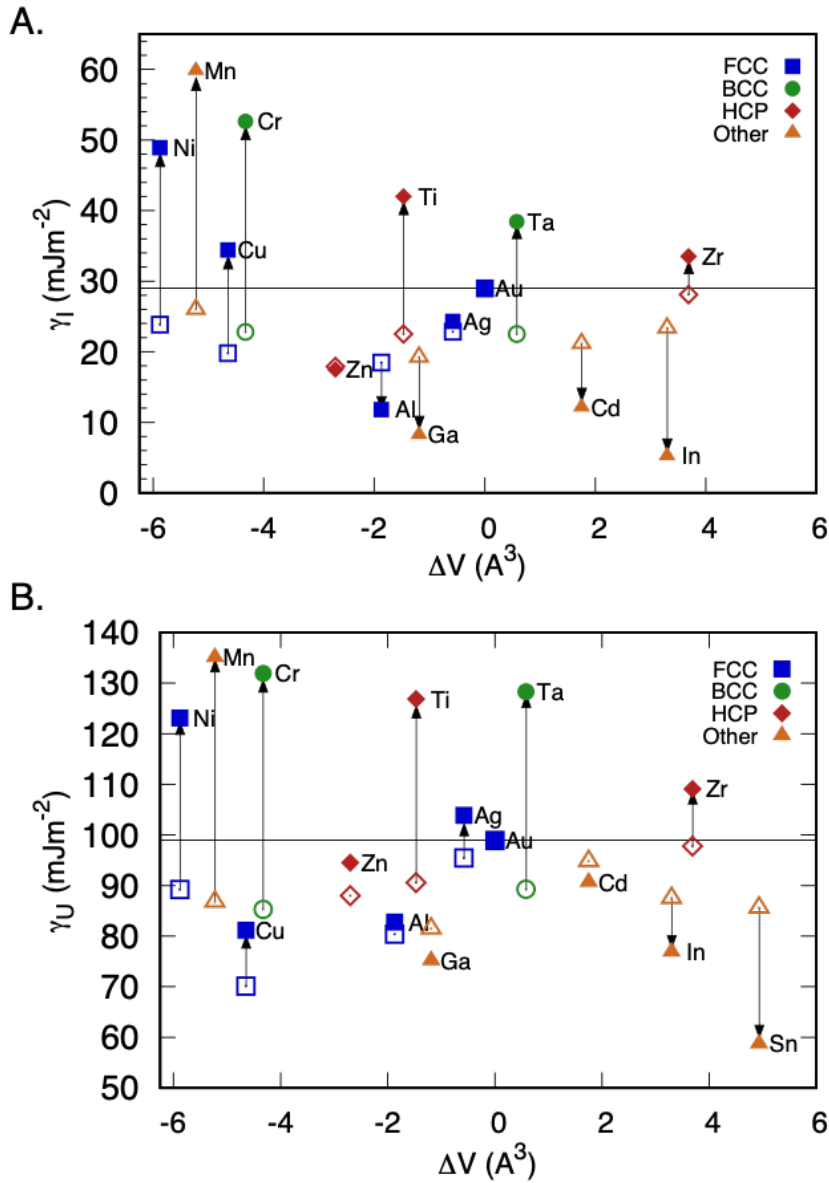

Figure 6: A. Intrinsic stacking fault energy and B. Unstable stacking fault energy as function of as function of the defect relaxation volume ΔV of impurity atom substituted at Au site in $Au_{47}X$. Open symbols are for the site distant from the stacking fault region (in atomic plane #3, Fig. 4); closed symbols are for the site close to the stacking fault region (in atomic plane #1, see Fig. 4).

### 3.2.2 Stacking fault energy as function of alloy concentration

The stacking fault energy in alloys systems is a strong function of alloy concentration.[70] The computed stacking fault energies for approximately 1%, 2% and 4% overall concentrations of the alloying elements are shown in Fig. 7. As described above, in our calculations we substitute single Au atom with an alloy atom in different supercell sizes to simulate 1%, 2% and 4% alloy compositions. This provides a unique choice for substituting a Au atom with an alloying atom because Au atoms are all identical in a given plane from the SF region. The alloy concentration in the specific (111) plane is much higher, typically 6.2%, 12%, and 25% compared to the respective overall concentrations. Therefore, any changes in the stacking fault energies due to substitution of alloying atom compared to pure Au should be significant. The results shown in Fig. 7 are the average stacking fault energies defined as

$$\gamma_{avg} = \frac{\sum_n \gamma_n}{\sum n}, \tag{5}$$

where $\gamma_n$ is the SFE of substituting the alloying atom in the nth (111) atomic plane with n = 1 to 6 and at a fixed distance from the stacking fault plane, as shown in Fig. 4. The decrease and increase in stable SFE (shown in Fig. 6) is consistent with alloy concentrations for all the alloying elements considered, with higher alloying concentration normally resulting in larger changes in the stacking fault energies.

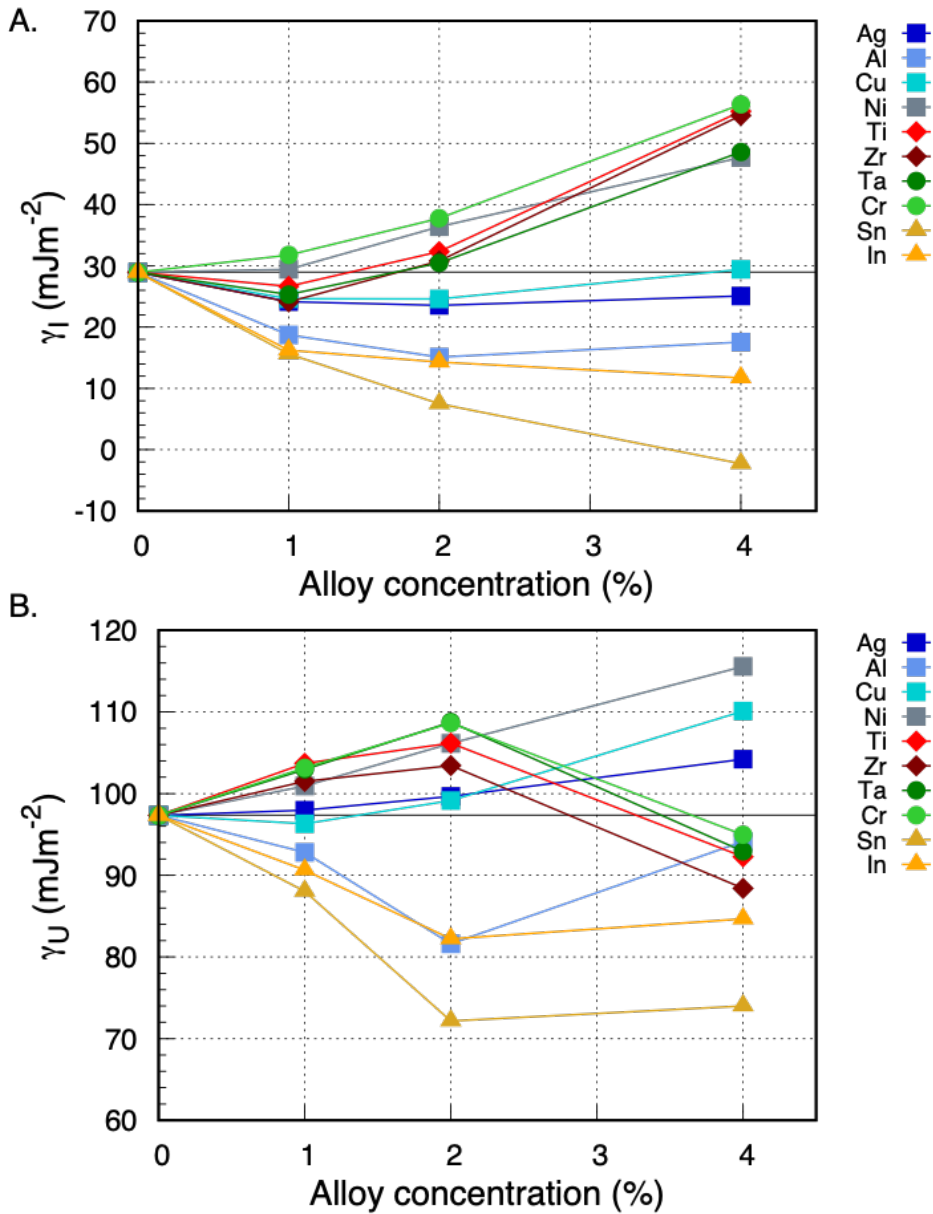

Figure 7: A. Average stable (or intrinsic) and B. unstable stacking fault energies for different FCC (squares), BCC (circles), HCP (diamonds) and other (triangle) alloying atoms as function of concentration. 1%, 2% and 4% alloy concentrations are simulated in $Au_{95}X$, $Au_{47}X$ and $Au_{23}X$ DFT supercells, respectively.

Al, Ag, Cu, Sn and In show decreases in stable stacking fault energies for all the investigated concentrations, whereas alloying elements Ti, Ni, Zr, Ta and Cr show increases in stable SFEs with increasing concentrations. Unstable SFEs for Ti, Zr, Cr and Ta show both increase and decrease with increasing alloy concentration; suggesting no direct dependence between the change in the unstable SFEs when compared to the change in the stable SFEs. We analyze changes in both stable and unstable SFEs and their dependence in the next section and discuss their implications to the deformation behavior in more detail. In a previous study, we have also performed SFE calculations on pure Au and all possible alloys using Embedded Atom Method (EAM) potentials[71]. In the EAM calculations, there are large numbers of substituted alloying elements, which are randomly distributed around the stacking fault region. Therefore, calculated stacking fault energies values are more likely to average out from the position dependence. We analyze and discuss some of the implication of these results to strengthening in Au alloys and separate indentation simulations[71] in the next section.

## 4. Discussion and analysis

### 4.1 Stacking fault energy of pure Au

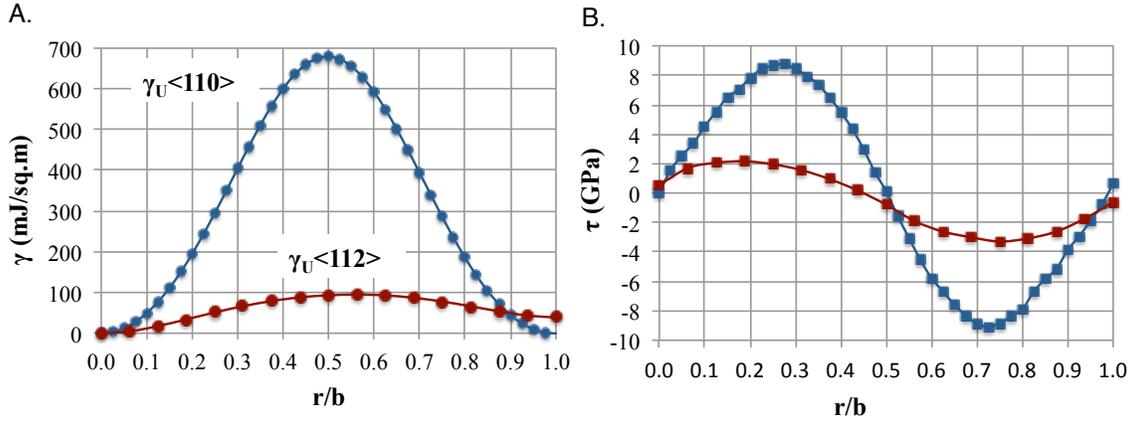

Figure 8: A. Generalized stacking fault energy curve (circles), and B. theoretical shear stress (squares) for slip systems {111}<112> (red) and {111}<110> (blue). The Burgers vector for the slip systems are $b_{\langle 112 \rangle} = \frac{1}{6}\langle 112 \rangle$ and $b_{\langle 110 \rangle} = \frac{1}{6}\langle 110 \rangle$, respectively. Lower $\gamma_U \langle 112 \rangle$ and $\tau_C \langle 112 \rangle$ indicates a more favorable {111}<112> slip system for plastic deformation.

Figure 8 compares the DFT calculated stacking fault energies and theoretical shear stresses during the shear deformation on (111) plane along the $[\bar{1}10]$ and $[11\bar{2}]$ directions, respectively. Based on the unstable SFEs, the energy barrier per unit area for atoms in (111) planes to shear along [1̄10] direction is 678 mJ/m² , which is much larger than the energy barrier of 102 mJ/m² needed to shear atoms along the [112̄] direction. Therefore, formation of partials in Au is energetically much more favorable compared to the glide of the full dislocation during the shear deformation. This is also consistent with the observation of stacking faults in indentation studies on Au thin films that have clearly shown plastic deformation occurring via a three-step mechanism involving nucleation, glide, and reaction of Shockley partials on {111} planes[8,32,71–74].

Calculation of stacking fault energies alone does not enable direct comparisons with the plastic deformation observed in the indentation studies, because $\gamma_U$ is a static quantity and may not be sufficient to describe the dynamics of slip in indentation experiments and simulations. A more relevant quantity is the theoretical shear stress $\left(\tau^{th}\langle\vec{r}\rangle = \frac{\partial \gamma}{\partial \vec{r}}\right)$ required to initiate and maintain slip along the slip direction[73]. The motion of atoms is along the direction with the smallest value of shear stress. This is commonly referred to as the critical resolved shear stress (CRSS), $\tau_C$. From Fig. 8B, the $\tau_C$ for {111}<112> slip system is 2.16 GPa, which is in good agreement with experimental estimates[8] of 1.5-2.0 GPa and other theoretical results[73]. Thus, our DFT calculations can explain why 1/6 <112> partial dislocations are favored during plastic deformation in pure Au, over other possible ½<110> deformation modes, which is in line with previous theoretical and indentation studies in pure Au.[73] Additionally, issues such as the deformation behavior changes in single crystal Au with dilute alloy addition, have not been addressed to this point, and thus motivate the next section of our study.

4.2    Stacking fault energy of Au alloys

From our simulation results on stacking faults in Au alloys in Sec. 3.2.1, we observe a strong dependence of the SFE on the position of the alloying element with respect to the stacking fault layer, as well as on the concentration of the alloying elements. To further understand the nature of this dependence, we explore the relation between the compositional dependence of SFE on the volume change associated with alloying elements. The changes in SFE when substituting an alloying element close to the fault versus away from the fault layer, is analogous to the concept of the segregation energy[75], familiar in the context of surfaces, grain boundaries[76] and dislocations[77]. As mentioned above, for stacking faults this is often referred to in the literature as Suzuki segregation[37]. In our calculations we define this change in stacking fault energy $\Delta \gamma$ as the difference between the total energy of the system with alloying element at the stacking fault layer and the total energy when alloying element is placed in a layer farthest away (layer #3 in Fig. 4) from the stacking fault. The calculated $\Delta \gamma$ as function of the relaxation volume $\Delta V$ of the alloying element is shown in Fig. 9.

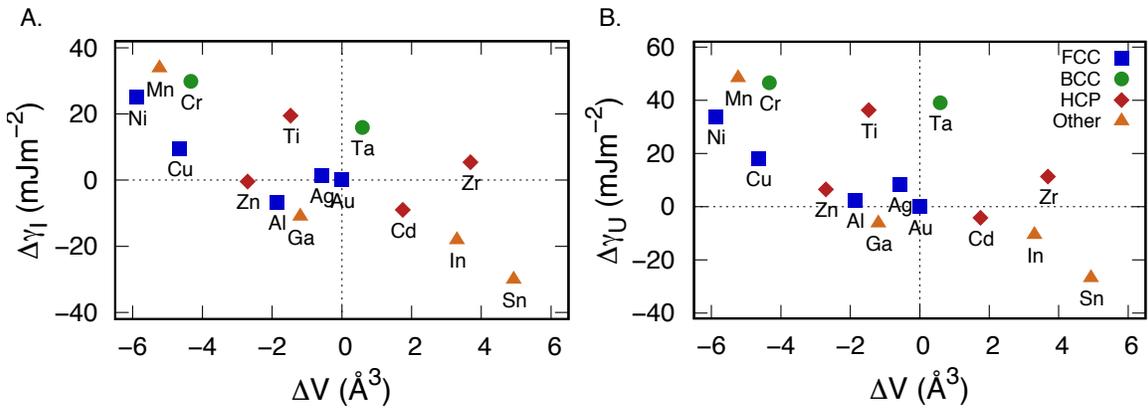

Figure 9: A. Change (Energy close to SF region – Energy farthest away from the SF region) in intrinsic stacking fault energy, and B. Change in the unstable stacking fault energy as a function of relaxation volume, i.e., the change in volume when an impurity atom is substituted at Au site in $Au_{47}X$ simulation cell.

The change in the stacking fault energy for a given alloying element in Au decreases approximately linearly with increasing relaxation volume (see Fig. 9) and is independent of the type of crystal structure of the alloying element. Similar trends in change in SFE with relaxation volume of alloying element are also observed in a recent study on dilute Ni alloys[40]. The larger the size mismatch between the alloying element and Au, the larger is the change in the stacking fault energy. Alloying elements such as Ni, Ti, Mn, Cr, Cu and Ag with smaller volumes than Au tend to substitute away from the stacking fault, whereas, Sn, In and Cd with larger volumes tend to segregate at the stacking fault layer. This change in stacking fault energy with relaxation volume when an alloying element is placed in the stacking fault layer suggests that the energetics of the process depend upon the interaction of the local misfit due to the substituted alloy atom with the strain at the stacking fault, analogous to segregation of impurities to dislocations[77,78]. However, as shown in Fig. 9, there are elements such as Zr, Ta, Ga and Al for which the change in SFE deviates from the general trend based on the change in the relaxation volume (or the size mismatch). This seems to indicate that electronic and chemical factors also contribute to interaction of alloying elements with the stacking fault.

In order to make better predictions of the deformation mechanism and ultimately the hardening/softening behavior in the alloy, we compare energy barrier $(\gamma_U - \gamma_I)$ and the ratio $\gamma_I/\gamma_U$ obtained from the GSFE curves for all the investigated alloying systems in Fig. 9. It has been suggested[11,19] that in addition to $\gamma_I$, $\gamma_U$ should also be taken into consideration while describing processes involving nucleation and slip of partial dislocations. Dislocation nucleation[11] depends on the barrier height $(\gamma_U - \gamma_I)$ and slip on the width of the stacking fault regions (separation between partial dislocation) that varies inversely with $\gamma_I/\gamma_U$.[19] From Fig. 10 we broadly classify alloying elements into two groups:

Group 1: Ag, Cu, Al, Zn, Cd, Ga, In, and Sn have much smaller $\gamma_I/\gamma_U$ and the energy barrier $(\gamma_U - \gamma_I)$ is generally larger in these alloy systems than in pure Au.

Group 2: Cr, Ti, Zr, Ta, Mn and Ni have much larger $\gamma_I/\gamma_U$ and smaller barrier height $(\gamma_U - \gamma_I)$ than in pure Au.

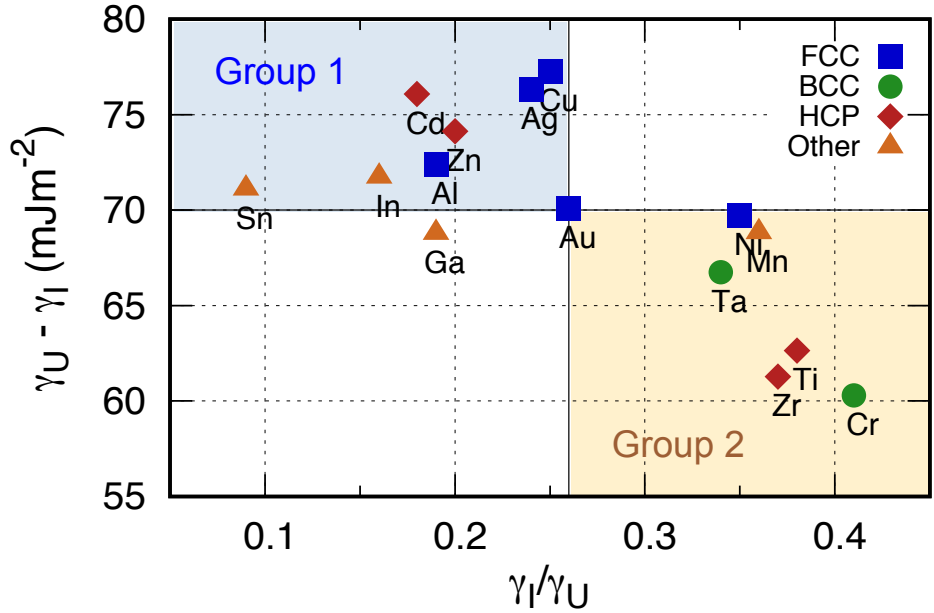

Figure 10: Difference in the unstable and stable SFE with respect to their ratio for various alloying elements in Au, calculated using DFT simulations. The $\gamma_I$ and $\gamma_U$ values are averaged over 1% and 2% compositions for each alloying element. Blue and yellow regions denote group 1 and group 2, respectively.

For group 1 alloys, SFEs are either smaller or very close in value to pure Au, but the relative decrease in $\gamma_I$ and $\gamma_U$ varies from element to element. Sn, In, Ga show a significant reduction in both $\gamma_I$ and $\gamma_U$, resulting in similar barrier height $(\gamma_U - \gamma_I)$ as in pure Au. However, for Ag and Cu $\gamma_I$ decreases while $\gamma_U$ increases marginally, keeping the ratio close to that of pure Au, but with a larger barrier height. For group 2 alloys, SFEs are generally larger than pure Au, but with a relatively larger increase in $\gamma_I$ than in $\gamma_U$. This suggests that $\gamma_I$ and $\gamma_U$ can increase or decrease independently, making it hard to predict change in $(\gamma_U - \gamma_I)$ by just examining the change in either $\gamma_I$ or $\gamma_U$.

Though the actual mechanical behavior in metallic alloys depends on microstructure and experimental conditions, it is underpinned by the SFEs. Here we highlight some of the key observations about the deformation behavior in metallic systems. SFs involve the formation of the leading partial dislocation that needs to overcome an energy barrier of $\gamma_U$, while the trailing partial need to overcome the barrier of $(\gamma_U - \gamma_I)$. The larger value of $\gamma_U$ or $(\gamma_U - \gamma_I)$, effectively limits the nucleation of partial dislocations and therefore contributes to an increase in the hardness of the material. Additionally, the width of the stacking fault affects the ease of cross slip[24]. Metals such as Au and Cu with much lower $\gamma_I$ and $\gamma_I/\gamma_U$ values and larger SF width, are more like to show hardening due to the difficulty in cross-slip, compared to metals like Al with larger $\gamma_I$ and $\gamma_I/\gamma_U$ and smaller stacking fault width.

Group 2 alloying elements (Cr, Ti, Zr, Ta, Mn and Ni), which have a higher $\gamma_I/\gamma_U$ ratio compared to pure Au are more likely to result in a smaller SF width than pure Au. Dislocation nucleation in these systems is likely to be easier due to the lower barrier height and hence, these elements are more likely to result in softening when alloyed with Au. However, group 1 alloying elements (Ag, Cu, Al, Zn, Cd, Ga, In, and Sn) are likely to result in larger stacking fault width and higher barrier to dislocation nucleation. Therefore, based on the difficulty in dislocation nucleation and cross-slip, these elements are more likely to show solid solution strengthening in pure Au. An experimental investigation by Jax et al.[7] on dilute (0.8 – 1.6 %) binary Au alloys reports solid solution based hardening in Au-In,-Cd,-Zn,-Ga alloys, which is in consistent with our analysis. Jax and co-authors observed that while size misfit of alloying element correlates with hardness, it is not sufficient to explain the relative hardness trends within the explored binary alloys. In an earlier nanoindentation study[71], we showed that the deformation behavior in single crystal Au alloy systems depends critically on SFE, as it control the nucleation and interaction of partial dislocations in these systems. From these nanoindentation simulations, we found that Ag, Cu and Al lead to strengthening, whereas, Ti leads to softening; these results are consistent with the qualitative prediction based on our DFT calculations of SFEs in alloy systems.

## 5. Conclusions

Using first-principles based DFT calculations we have systematically explored the influence of dilute concentrations of various alloying elements on the SFE of pure Au. We demonstrated that both concentration and position of the alloying element with respect to the stacking fault region has a strong influence on the stable and unstable SFE in Au alloys systems. Changes in the SFEs are found to be strongly dependent on the relaxation volume introduced by the alloying element. Further, based on the generalized SFE curves, we predict that elements such as Ag, Cu, Al, Zn, Cd, Ga, In and Sn are likely to show solid solution strengthening in pure Au. These findings are in good agreement with our recent molecular dynamics nanoindendation simulations on single crystal Au alloys systems as well as with previously available experimental data in Au alloys.

**Data availability**

The raw/processed data required to reproduce these finding will be made available on reasonable request.

**Acknowledgements**

This work of AG, YL, AC and SRP was supported by Seagate Technology. The work of SBS was supported by Air Force Office of Scientific Research through Grant # FA9550-12-1-0456.